\documentclass[aps,prl, twocolumn,superscriptaddress,floatfix]{revtex4}
\usepackage{amsmath,mathtools, amssymb,graphicx,color,overpic,float}
\begin{document}

\title{Single-Temperature Quantum Engine Without Feedback Control}
\author{Juyeon Yi}
\affiliation{Department of Physics, Pusan National University, Busan 609-735}
\author{Peter Talkner\footnote{corresponding author:peter.talkner@physik.uni-augsburg.de }}
\affiliation{Institut f\"ur Physik, Universit\"at Augsburg, Universit\"atsstra{\ss}e 1, 86159 Augsburg, Germany}
\author{Yong Woon Kim\footnote{corresponding author: y.w.kim@kaist.ac.kr}}
\affiliation{Graduate School of Nanoscience and Technology, Korea Advanced Institute of Science and Technology, Daejeon 305-701, Korea}

\date{\today}

\hspace{2cm} \\

\begin{abstract}
A cyclically working quantum mechanical engine that operates at a single temperature is proposed.
Its energy input is delivered by a quantum measurement. The functioning of the engine does not require any feedback control.   
We analyze work, heat, and the efficiency of the engine for the case of a working substance that is governed by the laws of quantum mechanics and that can be adiabatically compressed and dilated. The obtained general expressions are exemplified for a spin in an adiabatically changing  magnetic field and a particle moving in a potential with slowly changing shape. 

\end{abstract}

\maketitle

An engine converts 
some form of energy into mechanical work in a cyclic process that can be repeated at whim. An important example  
is a heat engine designed to 
utilize the energy exchange between
heat reservoirs at different temperatures~\cite{landau,planck,schroedinger}. The  research 
of heat engines has a long standing history, going back to
the industrial revolution in the $18$th century. 
Recently, motivated by Feynman's quote ``There's plenty of room at the bottom''~\cite{Feynman},  
the interest in the working principles of engines functioning on mesoscopic and also on  molecular and atomic scales, as well as in their designs and optimization has grown substantially. 
A major challenge for the understanding of small scale engines is to cope with the presence of unavoidable random noise in the form of thermal fluctuations \cite{Reimann,Hanggi,seifert}. At sufficiently low temperatures, quantum effects such as coherence and quantum noise may also become relevant. Related questions that have been discussed in the literature range from whether quantum effects are of any influence on an the performance of an engine at all, to whether quantum effects deteriorate or might even improve the performance of an engine~\cite{scovil,alicki,kosloff,bender, scully, kieu,koslofflevy,zheng,watanabe,marchegiani}.

 According to the second law of thermodynamics, an energy conversion of heat into work cannot be perfectly efficient~\cite{planck}. 
A finite amount of unconsumed energy must be dissipated into a low-temperature heat reservoir in order to restore the initial state and 
complete a cyclic process. Therefore, heat engines operating at a single temperature do not exist. This is also a significant consequence of the fluctuation theorem~\cite{jarzynski, tasaki} stating that for a cyclic process with only one temperature involved, the average of work {\it done on} the system $\langle W\rangle$ cannot be negative~(see, for a review, \cite{campisi}).
 
An engine operating at a single temperature does yet exist if one allows a Maxwell demon to help, or, in other words if a feedback control is active as in a Szilard engine~\cite{szilard, leff}.  The seeming contradiction to the second law 
can be resolved if the feedback control mechanism is included into the dynamics of the system constituting the engine \cite{stengel}. 
On a more formal level it can be understood in terms of the information gain $I >0$  as the result of a measurement which is part of the feedback control~\cite{touchette}. 
The information gain leads to a modified, negative lower bound of the average work done by the system as $\langle W \rangle \geq - k_{B}T I$~\cite{kawai, sagawa2}. This idea has been applied to classical~\cite{berut,toyabe,gavrilov} and quantum systems~\cite{zurek,kim,dong}.  

In this Letter, we propose a cyclically working quantum engine at a single temperature without feedback control. 
An essential ingredient of the engine protocol is a quantum measurement performed on the working substance of the engine. The result of this measurement is ignored and therefore cannot trigger any control. This measurement would be ineffective for an engine working according to the laws of classical physics. 
In a quantum system, however, a measurement imposes a change of the state and consequently an increase of the energy of the system which is the working substance in the present context.
The working substance is described by a Hamiltonian $H(\lambda)$ depending on a parameter $\lambda$. At the beginning of any cycle the parameter assumes the value $\lambda_i$ and the working substance is in a canonical equilibrium  state at the temperature $T$ ($k_{B}T=\beta^{-1}$). Starting with this state {\bf 0} a cycle consists of two adiabatic processes interrupted by the measurement and a final thermalization step. This series of ``strokes'' is sketched as follows:
\begin{equation}
{\bf 0} \xRightarrow{\text{AP I\; }} {\bf 1} \xRightarrow{\text{QM}} {\bf 2} 
\xRightarrow{\text{AP II}} {\bf 3}
\xRightarrow{\text{T\;}}{\bf 0}\:.
\label{0123}
\end{equation}

{\it Adiabatic process I}: The first stroke AP\:I is an adiabatic compression caused by a sufficiently slow parameter change from $\lambda_i$ to $\lambda_f$. 
Under the assumption that the energy levels of the Hamiltonian $H(\lambda)$ do nowhere cross as a function of the parameter $\lambda$, in this stroke the occupation probabilities of the energy branches do not change~\cite{born}.  
This stroke is required to lead to an increasing level spacing. In this sense it corresponds to a compression.

{\it Quantum measurement}: While keeping the Hamiltonian $H(\lambda_{f})$ fixed, a quantum measurement (QM) of an observable that does not commute with $H(\lambda_f)$ is performed on the working substance. The measurement causes a state change which will be considered as instantaneous. This state change implies a change of the occupation probabilities of the energy eigenstates.

{\it Adiabatic process II}:  Subsequent to the measurement stroke QM, during AP\:II the system undergoes an adiabatic expansion by means of a slow parameter change from $\lambda_f$ back to $\lambda_i$. During this stroke the occupation probabilities of the energy levels stay constant at those values reached immediately after the measurement.

{\it Thermalization}: Finally, while the parameter is kept at $\lambda_f$ the system is brought into a weak contact with a thermal reservoir at the initial temperature $T$. After a sufficiently long time has elapsed the system again reaches the initial equilibrium state, and the cycle can be resumed with the adiabatic process I.

In any thermodynamic process, the energy change of a system can be decomposed into work and heat. A change of energy is considered as work if it is caused by the variation of an externally controlled parameter ($\lambda$ in our case) and as heat if it results from a contact of the system with its environment.
We adopt the convention to consider both heat and work as positive if the energy of the working substance increases. 
According to this definition the energy changes caused by the adiabatic processes AP\:I and AP\:II must be classified as work while in the other two strokes QM and T heat is exchanged. For the measurement stroke a measurement apparatus and for the thermalization a heat bath represent the respective environments  with which energy is exchanged.  
We now demonstrate that work can be done within a complete cycle as described above.

{\it Work done during the adiabatic process I}:
During the stroke AP\:I the working substance is thermally isolated and its dynamics is governed by a slowly changing Hamiltonian $H(\lambda) = \sum_n E_n(\lambda) |n;\lambda\rangle \langle n; \lambda|${\color{blue},} where $E_n(\lambda)$ and $| n;\lambda \rangle$ are the corresponding eigenvalues and eigenstates, respectively. Here we assume that the eigenvalues $E_n(\lambda)$ are not degenerate for all considered values of $\lambda$.      
Further, by assumption, the working substance is initially staying in 
a canonical equilibrium state. Hence, the
population of its energy levels is determined by 
\begin{equation}\label{gibbs}
p^{eq}_{n}(\lambda_{i})=e^{-\beta E_{n}(\lambda_{i})}/Z,
\end{equation}
where $Z=\sum_{n}e^{-\beta E_{n}(\lambda_{i})}$ is the
canonical partition function. 
In the course of the adiabatic stroke the occupation probabilities of the energy eigenstates remain unchanged at $p^{eq}_{n}(\lambda_{i})$ such that the state of the working substance at the time when the parameter $\lambda$ is reached is given by the following density matrix  
\begin{equation}
\rho_{I}(\lambda)= \sum_{n}p^{eq}_{n}(\lambda_{i})|n;\lambda \rangle \langle n;\lambda |~.
\end{equation}
The work in this process is given by the differences of the energies $E_n(\lambda_f) -E_n(\lambda_i)$ occurring with the
probability $p^{eq}_n(\lambda_i)$. Consequently, the average work $W_I$ in the stroke AP\:I becomes  
\begin{equation}\label{w1}
W_{I}=\sum_{n}\left [E_{n}(\lambda_{f})-E_{n}(\lambda_{i}) \right ]p_{n}^{eq}(\lambda_{i})~. 
\end{equation}
In passing we note that, in contrast to the specification of work done in an arbitrary protocol requiring two energy measurements, here, due to the adiabaticity of the stroke, a single energy measurement, which could be performed at any instant of time during the stroke, suffices.   

{\it Post-measurement state}: In the next stroke a measurement of an observable $A$ with the eigenvalues $a_\alpha$ as possible results is performed. We consider the class of minimally disturbing generalized measurements which can be characterized by hermitian measurement operators $M_\alpha= M^\dagger_\alpha$ satisfying $\sum_\alpha M^2_\alpha = \openone$~\cite{wiseman}.
In a non-selective measurement  the post-measurement state assumes then the form
\begin{equation}
\rho_{PM}=\sum_{\alpha}M_{\alpha}\rho_{I}(\lambda_{f})M_{\alpha}\:.
\label{rpm}
\end{equation}
For $M_\alpha$ agreeing with the projection operators onto the eigenspaces of the observable $A$, the standard result of a projective measurement is recovered, but more general measurement schemes can be described in this way.
From the form (\ref{rpm}) of the post-measurement state, one obtains the expression $p(m,n) = T_{m,n} p^{eq}_n(\lambda_i)$ for the joint probability $p(m,n)$ of finding the eigenstate with label $n$ before and the one with label $m$ after the measurement. Here the transition probability $T_{m,n}$ is given by
 \begin{equation}\label{tmn}
T_{m,n}\equiv\sum_{\alpha} |\langle n;\lambda_{f}|M_{\alpha} |m;\lambda_{f}\rangle|^{2}~.
\end{equation}  
Hence, the average energy change $Q_M$ of the working substance caused by the measurement  becomes
\begin{equation}
\begin{split}
\label{me}
Q_M&=\sum_{m,n} \left [E_m(\lambda_f) - E_n(\lambda_f) \right ] T_{m,n} p^{eq}_n(\lambda_i)\\
&=\frac{1}{2} \sum_{m,n} \left [ E_m(\lambda_f) - E_n(\lambda_f) \right ] T_{m,n}\\
&\quad \times \left (p^{eq}_n(\lambda_i) - p^{eq}_m(\lambda_i) \right ) \geq 0 \:,
\end{split}
\end{equation}
where the second equality is implied by the following properties of the transition matrix \cite{md},
\begin{equation}\label{tranprop}
\sum_{m}T_{n,m}=1,~~T_{n,m}=T_{m,n}\:. 
\end{equation}
The expression of the second line in Eq. (\ref{me}) cannot become negative because $T_{n,m} \geq 0$ 
and the probabilities $p^{eq}_n(\lambda_i)$ decrease with increasing energies $E_n(\lambda_f)$. Therefore the amount of heat transferred from the measurement apparatus to the working system is always positive. In a way it acts as the hot reservoir of a heat engine. 
Using the expression (\ref{tmn}) for the transition probability, one can write the measurement heat $Q_M$ as
\begin{equation}
Q_M= \sum_n \langle n;\lambda_f| H_M(\lambda_f) -H(\lambda_f)|n; \lambda_f \rangle p^{eq}_n(\lambda_i) \:,
\label{QMH}
\end{equation}
where
$H_M(\lambda_{f}) = \sum_\alpha M^\dagger_\alpha H(\lambda_{f}) M_\alpha $.
For measurement operators $M_\alpha$ commuting with $H(\lambda_f)$ one finds $H_M(\lambda_f) = H(\lambda_f)$, and consequently,  the energy supplied by the measurement is only different from zero if the measurement operators $M_\alpha$ do not commute with the Hamiltonian.  

{\it Work done during the adiabatic process II}: 
The second adiabatic stroke AP\:II reverts the first one in changing the parameter from $\lambda_f$ back to the initial value $\lambda_i$. Analogously to the argument leading to Eq. (\ref{w1}) the work $W_{II}$ done by the working substance is given by
\begin{equation}\label{w2}
W_{II}=  \sum_{n}\left [E_{n}(\lambda_{i})-E_{n}(\lambda_{f}) \right ]p_{n}^{PM} \:,
\end{equation}
where $p_{n}^{PM}$ denotes the probability of finding the $n$th eigenstate in the post-measurement state (\ref{rpm}). It is given by
\begin{equation}
p_{n}^{PM} \equiv \langle n;\lambda | \rho_{PM}|n;\lambda\rangle = \sum_{m}p^{eq}_{m}(\lambda_{i})T_{m,n} \:.
\label{pnpm}
\end{equation}

The energy change of the working substance in the final stroke T  is caused by the contact with a heat bath at temperature $T$. Its average therefore is a heat, which we denote by $Q_T$.  
It can be expressed as
\begin{equation}
Q_T=\sum_{n}E_{n}(\lambda_{i}) \left [p_{n}^{eq}(\lambda_{i})-p_{n}^{PM} \right] \leq 0~.
\label{QT}
\end{equation}
Along the same lines of arguments leading to the positive sign of $Q_M$ one finds that  $Q_T$ is negative and hence energy is flowing from the working substance into the heat bath.

Finally, we determine the total average work $W$ done by the system as the sum of $W_{I}$ and $W_{II}$ which are given by the Eqs.~(\ref{w1}) and (\ref{w2}), respectively.
This sum can be expressed as
\begin{equation}\label{totw2}
W=\frac{1}{2}\sum_{n,m}(\Delta^f_{m,n}-\Delta^i_{m,n})T_{m,n}\left [p_{m}^{eq}(\lambda_{i})-p_{n}^{eq}(\lambda_{i}) \right ]\:,
\end{equation}
where $\Delta^\alpha_{m,n}$ denotes  the level distance between the $m$th and the $n$th energy eigenvalues of the Hamiltonian $H(\lambda_\alpha)$ for $\alpha = i,f$ as given by  
\begin{equation}
\Delta^\alpha_{m,n} \equiv E_m(\lambda_\alpha) - E_n(\lambda_\alpha) \quad \alpha = i,f \:. 
\label{envari}
\end{equation}
In order to determine the sign of the total work we separately consider pairs of $n$ and $m$ leading to different signs of $\Delta^i_{n,m}$. 
If $\Delta^i_{m,n} >0$, 
then the level-distance grows because of the compression in going from $\lambda_i$ to $\lambda_f$, and hence $\Delta^f_{m,n} \geq \Delta^i_{m,n}$. Because of the 
monotonic 
decrease of the canonical probability $p^{eq}_k(\lambda_i)$ with increasing energy $E_k(\lambda_i)$, the difference $p^{eq}_m(\lambda_i) - p^{eq}_n(\lambda_i)$ is negative; taking into account the positivity of the transition probabilities $T_{m,n}$, all contributions to the right hand side of Eq. (\ref{totw2}) with $\Delta^i_{m,n} >0$ are negative. Similarly, $\Delta^i_{m,n} \leq 0$ implies $\Delta^f_{m,n} \leq \Delta^i_{m,n}$ and $p^{eq}_m(\lambda_i) - p^{eq}_n(\lambda_i)\geq 0$, also leading with $T_{m,n}\geq 0$ to a non-positive contribution to the total work. 

Summarizing we note that within a cycle as sketched in (\ref{0123}) part of the energy $Q_M$ injected by a non-selective measurement can be extracted as work $W$ by means of adiabatic processes. The remaining energy $Q_T$ is dumped as heat into a reservoir at the temperature $T$. The efficiency of the engine is given by
\begin{equation}
\eta = \frac{-W}{Q_M} = 1 - \frac{\sum_{m,n} \Delta^i_{m,n} T_{m,n} p^{eq}_n(\lambda_i)} {\sum_{m,n} \Delta^f_{m,n} T_{m,n} p^{eq}_n(\lambda_i)} \:.
\label{eta}
\end{equation}
In the particular case of uniform compression described by $\Delta^i_{m,n} = \gamma \Delta^f_{m,n}$ with $\gamma$ less than one and independent of $m$ and $n$ the efficiency becomes $\eta = 1-\gamma$.

We expect that such an engine will be characterized by a smaller but still positive efficiency if the parameter $\lambda$ is varied at a finite speed rather than adiabatically.  A detailed discussion of this issue will be presented elsewhere. 
  
{\it Examples.} We illustrate our findings by two specific examples. In the first one we choose a spin 1/2 in an external magnetic field as the working substance, which hence is governed by the Hamiltonian $H(B) = - \mu_B B \sigma_z$, where $\mu_B$ is the Bohr magneton and $\sigma_z$  the $z$ component of the Pauli spin matrix. The magnetic field $B$, which is supposed to point in  the $z$-direction, plays the role of the external parameter $\lambda$ changing in the AP\:I stroke from $B_0>0$ to $B_1>B_0$ and later in the AP\:II stroke back again to $B_0$. The energy eigenvalues of $H(B)$ are $E_\pm(B) = \mp \mu_B B$ in the spin-up (+) and the spin-down (-) state, respectively. The initial populations of these states are given by the canonical probabilities $p^{eq}_{\pm}(B_0) = e^{\pm \beta \mu_B B_0}/Z$,  
where the partition function is given by $Z= 2\cosh (\beta \mu_B B_0)$. The measurement stroke QM is done as a projective measurement of the spin-component $\sigma_x$. It is hence characterized by the measurement operators $M_\pm=(1\pm \sigma_x)/2$ 
yielding the transition probability $T_{\pm,\pm} =1/2$ between any pair of states  as well as uniform post-measurement probabilities $p^{PM}_\pm = 1/2$. Due to the uniform population of the energy eigenstates after the $\sigma_x$ measurement, the work done in the AP\:II stroke vanishes, and the total work is given by that of AP\:I, which, with Eq. (\ref{w1}) yields
\begin{equation}
W= W_I= \mu_{B}(B_{0}-B_{1})\tanh(\beta \mu_{B}B_{0}) < 0~.
\end{equation}
The amount of heat supplied to the system in the measurement stroke follows from Eq.~(\ref{me}) to read
\begin{equation}
Q_M =   \mu_B B_1 \tanh (\beta \mu_B B_0)\:.
\label{QMS}
\end{equation}
From Eq. (\ref{QT}) the heat dumped to the thermal reservoir results as
\begin{equation}
Q_T = - \mu_B B_0 \tanh (\beta \mu_B B_0)\:.
\label{QTS}
\end{equation}
Finally, the efficiency is given by 
\begin{equation}
\eta = 1-\frac{B_0}{B_1}\:.
\label{eS}
\end{equation}

In the second example the working substance consists of a particle of mass $m$ moving in a one-dimensional potential $V(\hat{x},\lambda)$. Its Hamiltonian hence is given by 
\begin{equation}\label{ham}
H(\lambda)=\frac{{\hat p}^{2}}{2m}+V({\hat x};\lambda)~,
\end{equation}
where ${\hat p}$ and ${\hat x}$ are the momentum and the position operator, respectively. 
The form of the potential $V(x,\lambda)$ (with $x$ being an eigenvalue of ${\hat x}$) can be controlled by the parameter $\lambda$. We mention as examples a free particle in a box of linear size $\lambda$ which is described by $V_{\text{box}}(x,\lambda)=0$ for $x\in(0,\lambda)$ and $V_{\text{box}}(x,\lambda) = \infty$ for $x \notin (0,\lambda)$, and a particle in a harmonic potential $V_h(x,\lambda) = \lambda x^2/2$ with curvature $\lambda$. In both cases a change of $\lambda$ from $\lambda_i$ to $\lambda_f$ as performed in AP\:I corresponds to a uniform compression with the compression factor $\gamma_{\text{box}} = (\lambda_f/\lambda_i)^2$ for the particle in a box  and $\gamma_h= \lambda_i/\lambda_f$ for the harmonic potential, provided $\gamma <1$. 
In this example we specify the QM stroke as a Gaussian position measurement that is characterized by the hermitean measurement operator $M_\alpha =(2 \pi \sigma^2)^{-1/4} e^{-(\hat{x}-\alpha)^2/(4 \sigma^2)}$ where $\alpha$ is the measured position and $\sigma^2$ the  variance of the measurement apparatus characterizing its precision. Note that the measurement operators are properly normalized according to $\int_{-\infty}^\infty d \alpha  M^2_\alpha = \openone$ and that in the limit $\sigma^2 \to 0$ a projective position measurement is approached. Because 
$\alpha$ is a continuous variable, the summation in the normalization of $M_\alpha$ becomes an integral. 
In order to determine the energy input $Q_M$ caused by the measurement we consider the difference $H_M(\lambda_f) - H(\lambda_f)$ which enters the expression (\ref{QMH}) for $Q_M$. Using the normalization of the measurement operators it can be written as $H_M(\lambda_f)-H(\lambda) = \int d\alpha M_\alpha [H(\lambda_f), M_\alpha]$ where $[\cdot,\cdot]$ denotes the commutator. Because $M_\alpha$ is a function of the position operator $\hat{x}$ but not of the momentum, only the kinetic part of the Hamiltonian contributes. 
The resulting commutator can be evaluated for the Gaussian $M_\alpha$ and found, after some algebra, to yield $H_M(\lambda_f) -H(\lambda_f) = \hbar^2/ (8 m \sigma^2)$. With Eq. (\ref{QMH}), we reach the expression for the energy delivered by the measurement,
\begin{equation}
Q_M = \frac{\hbar^2}{8 m \sigma^2} \:.
\label{QP}
\end{equation}
It is a remarkable fact that this result is independent of any detail of the potential and also independent of the temperature of the initial state of the working substance. It only depends on the mass of the particle and the variance of the Gaussian position measurement apparatus. It diverges in the limit of a projective measurement.
For a uniform compression the total work $W$ results as
\begin{equation}
W = - (1-\gamma) \frac{\hbar^2}{8 m \sigma^2}\:,
\label{WP}
\end{equation}
because then the efficiency is given by $\eta = 1-\gamma$.

{\it Conclusions.} We demonstrated that a non-selective measurement of any observable that does not commute with the Hamiltonian governing the dynamics of the system at the time of the measurement increases the energy of a quantum system. Because this energy gain is caused by the contact with a measurement apparatus, which itself is a quantum system, it can be counted as heat. This is in accordance with earlier observations for particular model systems that repeated measurements may heat up the system to reach infinite temperature \cite{schulman, erez, talkner,yi}. The amount of energy $Q_M$ delivered in a single measurement depends on the so-called operation $\phi$, characterizing the post-measurement state $\rho^{PM} = \phi(\rho) =\sum_\alpha M_\alpha \rho M_\alpha$ written in terms of the normalized measurement operators $M_\alpha$. It can be expressed as the difference of the energy average in the post-measurement state and the state $\rho$ immediately before the measurement yielding
\begin{equation}
Q_M = \text{Tr} H \phi(\rho) - \text{Tr} H \rho\:,    
\label{QMHH}
\end{equation}
see also Eq.~(\ref{QMH}). 
The positivity of the injected energy is a consequence of the symmetry of the transition matrix imposed by a minimally disturbing measurement and the decay of the Boltzmann weights with increasing energy.
 
 
Here we analyzed a cyclic process which works similarly as a heat engine with the only difference that the hot heat bath is replaced by a measurement.     
For the work strokes adiabatic compression and expansion processes are considered. No feedback mechanism is implemented. We found that the total work is negative, meaning that a part of the heat delivered by the measurement can be extracted as work. In general, the efficiency of such an engine as given by Eq. (\ref{eta}) depends on the temperature of the heat bath and the details of the eigenenergies in the initial dilated and in the final compressed state. For a uniform compression the efficiency simplifies to a mere function of the compression factor.

For the working substance any quantum system can be employed that can be compressed and dilated in terms of an externally controllable parameter $\lambda$. As special examples we considered a spin 1/2 in an external magnetic field that works as the controllable parameter and a particle in a deformable potential.  

In the present letter we considered only the averages of  work and heat. For the full understanding of the proposed type of engines the full statistics of heat and work caused by the thermal fluctuations of the heat bath and by the intrinsic quantum nature of the working substance will be relevant.    

{\it Acknowledgments.} This work was supported by the Deutsche Forschungsgemeinschaft via the projects HA 1517/35-1
and DE 1889/1-1 and also by Basic Science Research Program through the National Research Foundation of Korea(NRF) funded by the Ministry of Education, Science and Technology(Grant No. NRF-2013R1A1A2013137).


\begin{thebibliography}{10}
\bibitem{md}  Note that the symmetry of the transition matrix is a consequence of the assumption that the measurements are minimally disturbing meaning that $M_\alpha = M^\dagger_\alpha$. For other types of measurements the positivity of the energy change caused by the measurement need not be positive.
\end{thebibliography}
\end{document}